\documentclass[aps,twocolumn,floatfix,byrevtex,titlepage,balancelastpage,raggedbottom,amsfonts,amssymb,amsmath,showkeys,showpacs]{revtex4}
\usepackage{epsfig}
\usepackage{bm}
\usepackage{times}
\usepackage{mathptmx}
\usepackage{amsmath}
\usepackage{graphicx}
\usepackage{graphics}
\usepackage{amssymb}
\usepackage{enumerate}
\usepackage{bm}
\usepackage{longtable}
\usepackage{multirow}
\usepackage[english]{babel}
\begin{document}
%=======================================================================
\title{Wetting and contact-line effects for spherical and cylindrical droplets on graphene layers: A comparative molecular-dynamics investigation}
\author{Giulio Scocchi}
\email[To whom correspondence should be addressed: ]{giulio.scocchi@icimsi.ch}
\affiliation{University of Applied Sciences (SUPSI),
The iCIMSI Research Institute,
Galleria 2, CH-6928 Manno, Switzerland}
\author{Danilo Sergi}
\affiliation{University of Applied Sciences (SUPSI),
The iCIMSI Research Institute,
Galleria 2, CH-6928 Manno, Switzerland}
\author{Claudio D'Angelo}
\affiliation{University of Applied Sciences (SUPSI),
The iCIMSI Research Institute,
Galleria 2, CH-6928 Manno, Switzerland}
\author{Alberto Ortona}
\affiliation{University of Applied Sciences (SUPSI),
The iCIMSI Research Institute,
Galleria 2, CH-6928 Manno, Switzerland}
\date{\today}
%================================================
\begin{abstract}
In Molecular Dynamics (MD) simulations, interactions between water molecules and graphitic surfaces are often modeled
as a simple Lennard-Jones potential between oxygen and carbon atoms. A possible method for tuning this parameter consists
of simulating a water nanodroplet on a flat graphitic surface, measuring the equilibrium contact angle, extrapolating it
to the limit of a macroscopic droplet and finally matching this quantity to experimental results. Considering recent evidence
demonstrating that the contact angle of water on a graphitic plane is much higher than what was previously reported, we
estimate the oxygen-carbon interaction for the recent SPC/Fw water model. Results indicate a value of about
$0.2$ kJ/mol, much lower than previous estimations. We then perform simulations of cylindrical water filaments on graphitic
surfaces, in order to compare and correlate contact angles resulting from these two different systems. Results suggest that modified
Young's equation does not describe the relation between contact angle and drop size in the case of extremely small systems and that 
contributions different from the one deriving from contact line tension should be taken into account.  

\end{abstract}
%=======================================================================
% 47.11.Mn Molecular dynamics methods
% 47.55.D- Drops and bubbles
% 47.55.np Contact lines
%=======================================================================
%=======================================================================
\maketitle
%======================================================================

\section{Introduction}

Since the discovery of buckminsterfullerenes \cite{1} (awarded with the 1996 Nobel Prize in Chemistry) and carbon nanotubes \cite{2}, carbon
nanoparticles have been the subject of increasing scientific interest. In 2010, another Nobel Prize (this time in Physics) was assigned to Geim
and Novoselov \cite{3} for their research on graphene, the basic building block for graphitic materials \cite{4}, where graphitic materials are
defined as any allotropic form of carbon characterized by a six atom ring structure. Attention to this kind of materials is justified by their
peculiar properties as well as their possible applications, which range from electronics \cite{5} to polymer-based composite materials \cite{6}.
Amongst the different challenges presented by carbon nanoparticles processing, one of the most demanding consists of achieving fine dispersions of
these materials; this subject has indeed been thoroughly investigated in recent years, in experimental \cite{7,8,9,10,11} as well as in
modeling studies \cite{12,13,14,15,16}.
In simulation studies involving aqueous dispersions of graphitic materials, modeling the interactions between water molecules and carbon atoms
obviously represents one of the most important tasks. These interactions, regardless of the water model, are usually represented as a simple
Lennard-Jones potential between water oxygen and graphitic carbon atoms \cite{17,18,19,20,21,22,23,24,25,26,27,28,29}:
\begin{equation}
V(r)=4\varepsilon_{\mathrm{CO}}\Big[\Big(\frac{\sigma_{\mathrm{CO}}}{r_{\mathrm{CO}}}\Big)^{12}-\Big(
\frac{\sigma_{\mathrm{CO}}}{r_{\mathrm{CO}}}
\Big)^{6}\Big]\ ,
\label{eq:lj}
\end{equation}
where $\sigma_{\mathrm{CO}}$ determines the equilibrium distance, $\varepsilon_{\mathrm{CO}}$ is the depth of the potential well and $r_{\mathrm{CO}}$ is the
distance between a pair of oxygen and carbon atoms. Non-bond interaction between hydrogen and carbon are usually not accounted for, even if some
authors have investigated this option \cite{24}.
One way to proceed to estimate suitable values for these parameters consists of simulating a water droplet on a graphitic surface and comparing
the resulting equilibrium contact angle to experimental measurements, it being understood that the modified version of Young law for wetting
\cite{30,31,32}  has to be considered in order not to neglect the influence of contact line tension at the nanoscale \cite{young}:
\begin{equation}
\cos\theta=\cos\theta_{\infty}-\frac{\kappa}{\gamma r_{\mathrm{ca}}}\ .
\label{eq:young}
\end{equation}
In the above formula, $\theta$ and $\theta_{\infty}$ are the average actual and macroscopic contact angle respectively,
$\kappa$ is the contact line tension,
$\gamma$ is the superficial tension of water and $r_{\mathrm{ca}}$ is the contact area radius. In any case, it has to be noted that wetting
at the nanoscale still remains a largely unanswered question \cite{32}, to the point that it is not trivial to even define the concept of contact
angle in this size range \cite{33}. A similar uncertainty affects the definition and estimation of contact line tension, whose influence on contact
angle is considered significant only for droplets with a diameter ranging from one to some hundred nanometers, according to different authors
\cite{31,32}. Even the sign of line tension is not unambiguous and some authors even hypothesized it to be a mere artifact due to poor experimental
measurements, which are indeed known to be quite complex and sensitive \cite{32}. For a more detailed description of these issues, we refer to the
excellent reviews by M\a'endez-Vilas \textit{et al}.~\cite{32} and Amirfazli and Neumann \cite{34}.

In molecular dynamics studies, estimates of $\kappa$ were obtained by simulating nanodroplets of different sizes, plotting contact angles as a function
of droplet base radii (for a fixed set of Lennard-Jones parameters) and extending the relative linear fits for $1/r_{\mathrm{ca}}\rightarrow 0$. This
approach was followed by  Koumoutsakos and co-workers in an accurate series of papers \cite{18,24,25}, which constitutes a reference for MD
investigations of the same nature.
In their first study, along with a comprehensive review of water-carbon interaction potentials for different water models, the authors
provided a detailed description of the procedure (originally developed by Blake \textit{et al}.~\cite{35} and de Ruijter \textit{et al}.~\cite{36})
used for retrieving
the average contact angle from MD simulation of droplets on surfaces. Moreover, they found a direct linear relation between contact angle and water
monomer binding energy on graphite (actually modeled as two graphene sheets) and determined the values for $\varepsilon_{\mathrm{CO}}$ and
$\sigma_{\mathrm{CO}}$ which matched the experimental contact angle of water on graphite \cite{18}. In the following work, they explored the influence of
Lennard-Jones cutoffs and hydrogen-carbon interactions on contact angle  and compared a discrete graphitic plane model with a continuous one
\cite{24}. In the last paper of the series, they investigated the influence of fluid and surface impurities on the wetting properties of water on
graphite \cite{25}.

The values for $\varepsilon_{\mathrm{CO}}$ and $\sigma_{\mathrm{CO}}$ proposed in these papers \cite{18,24,25} have been
used for similar MD studies by many researchers \cite{14,19,21,22,23,24,25,26}. These two parameters have been adjusted in order to reproduce contact
angles of water on graphite ranging from $42^{\circ}$ to $86^{\circ}$, according to what was reported in the literature \cite{37,38,39,40} and generally
accepted by the scientific community \cite{41}. However, these experimental measurements are at least thirty years dated. Importantly, they were
performed
on graphite samples whose surface characteristics and purity, even in the case of pyrolytic graphite \cite{38} and highly-oriented pyrolytic graphite
\cite{39,40}, can hardly be compared to the ideal atomic smooth graphene sheets used in atomistic simulations to model graphitic surfaces \cite{svitova}.
In 2009, an experimental measurement closer  to the ideal conditions of the simulations reported above was performed by Wang \textit{et al}.~\cite{42}, who
produced samples consisting of few graphene layers and characterized them with WXRD, AFM and FTIR methods. In this study, the
contact angle of water on graphene layers turned out to be $127^{\circ}$.

%=======================================
\begin{figure}[b]
\includegraphics[width=8.5cm]{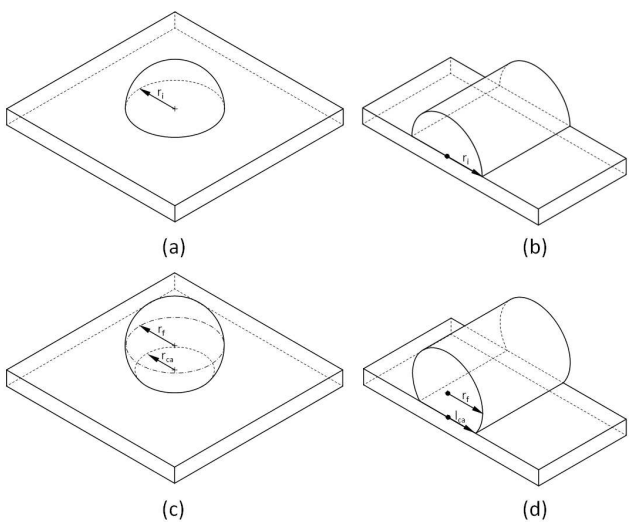}
\caption{
Schematic drawing of spherical droplets and cylindrical filaments in the initial configuration, (a) and (b), and
after equilibrium is reached, (c) and (d).
\label{fig:setup}}
\end{figure}
%=======================================
The measurements performed by Wang \textit{et al}.~are the most recent concerning water-graphite contact angle and in our opinion they
better reproduce the conditions represented in molecular models of such systems. Taking this into consideration, we decided to calculate
new Lennard-Jones
parameters for water-carbon interactions, using the recent flexible SPC/Fw water model \cite{43}. We basically followed the procedure reported by
Koumoutsakos and co-workers, although with some slight modifications which we will describe in the next section. Moreover, along with spherical
droplets, we also simulated cylindrical filaments \cite{44} (see Figure \ref{fig:setup}) of equivalent size and compared contact angle measurements
for these two different systems. Defining a correlation between results for spherical droplets and cylindrical filaments has two important outcomes:
first of all, it opens the possibility of avoiding time-consuming simulations of spherical droplets in further studies of static wetting properties of
water on graphitic surfaces, as already pointed out by other authors \cite{26}. Secondly, by analyzing the different influence of droplet and
filament size and shape on contact angle, it helps in understanding the molecular mechanisms determining nanoscale behavior of water on a graphitic
surface.
%======================================================================
%=======================================
\begin{table*}[t]
\begin{center}
\begin{tabular}{l|c|cc|cc}
\hline
\hline
  & this work & Wu \textit{et al.} & ($\%$ difference) & experimental  & ($\%$ difference)\\
\hline
$\langle r_{\mathrm{OO}}\rangle$ \AA & $1.0320$ & $1.03210$ & $0.1$ & $0.970$ & $6.4$\\
$\langle \theta_{\angle\mathrm{HOH}}\rangle$ & $107.47^{\circ}$ & $107.69^{\circ}$ & $0.2$ & $106^{\circ}$ & $1.4$\\
$\rho$ g/cm$^{3}$ & $1.025$ & $1.012$ & $1.0$ & $0.997$ & $2.8$\\
$D_{\mathrm{s}}$ $10^{-5}$cm$^{2}$s$^{-1}$ & $2.24$ & $2.32$ & $3.4$ & $2.3$ & $2.6$\\
\hline
\hline
\end{tabular}
\end{center}
\caption{\label{tab:bulk}
Bulk water properties: comparison and $\%$ difference with data from Wu \textit{et al}.~\cite{44} and experimental measurements
\cite{43}.}
\end{table*}
%=======================================
\section{Methodology}

In order to identify the most suitable value for $\varepsilon_{\mathrm{CO}}$ in the light of the recent experimental findings reported above, we followed
the procedure adopted by Koumoutsakos and co-workers, i.e. we performed a series of MD simulations of water droplets of different radii on two graphene
layers organized
as hexagonal graphite \cite{45} and varied the value of $\varepsilon_{\mathrm{CO}}$ in the range $0.1-0.3$ kJ/mol; then, we calculated the equilibrium
contact angle $\theta$ formed by the droplets on the surface and examined the relation between $\theta$ and $\varepsilon_{\mathrm{CO}}$. The value of
$\sigma_{\mathrm{CO}}$ was kept fixed at
$\sigma_{\mathrm{CO}}=3.190$ \AA$^{}$ \cite{18}. We chose to use the increasingly widespread SPC/Fw water model \cite{43,46,47}, for which the introduction
of
intramolecular degrees of freedom allows improved accuracy amongst the simple and common three point charge models. As previously said, both
cylindrical filaments and spherical droplets were simulated, in order to understand how to relate contact angle measurements resulting from these two
different systems. Model size was varied in order to assess its influence on droplet or filament shape above the graphitic surface and to allow
extrapolation of contact angles in the limit of very large base radius, i.e.~for macroscopic droplets. Finally, to confirm the behavior found using
spherical droplets with sizes up to $8'000$ molecules ($3'213$ for cylindrical filaments), we performed simulations of droplets composed of $16'000$ water
molecules ($5'117$ for the equivalent filament) using $\varepsilon_{\mathrm{CO}}=0.20$ kJ/mol. In the next sections, simulation and analysis details are
reported.  All computations were performed using Materials Studio 5.5 by Accelrys Software Inc.

%=======================================

\section{Simulations}

\textit{Water model.} We chose to base our study on the SPC/Fw water model. This model is strictly related to the widely
used Simple Point Charge (SPC) water model introduced by Berendsen and co-workers \cite{48}, the main difference consisting of bond and angle flexibility
within the single water molecule.  Considering the SPC/Fw model (with $r_{\mathrm{OH}_{1}}$, $r_{\mathrm{OH}_{2}}$ as the distance between the oxygen
and each of the two hydrogens and $\theta_{\angle\mathrm{HOH}}$  as the angle), the general interaction potentials can be expressed as \cite{43}:
\begin{multline}
V^{\mathrm{intra}}=\frac{k_{\mathrm{b}}}{2}[(r_{\mathrm{OH}_{1}}-r_{\mathrm{OH}}^{0})^{2}\\
+(r_{\mathrm{OH}_{2}}-r_{\mathrm{OH}}^{0})^{2}]+
\frac{k_{\mathrm{a}}}{2}(\theta_{\angle\mathrm{HOH}}-\theta_{\angle\mathrm{HOH}}^{0})^{2}
\end{multline}
\begin{equation}
V^{\mathrm{inter}}=\sum^{\textrm{all pairs}}_{i<j}
\Big\{4\varepsilon_{ij}\Big[\Big(\frac{\sigma_{ij}}{r_{ij}}\Big)^{12}-\Big(\frac{\sigma_{ij}}{r_{ij}}\Big)^{6}\Big]+\frac{q_{i}q_{j}}{r_{ij}}\Big\}\ ,
\end{equation}
where $V^{\mathrm{intra}}$ and $V^{\mathrm{inter}}$ account for bonded and non-bonded interactions, $r^{0}_{\mathrm{OH}_{1}}$  and $\theta^{0}_{\angle\mathrm{HOH}}$
are the equilibrium bond length and angle ($1.012$ \AA$^{}$ and $113.24^{\circ}$ respectively), $k_{\mathrm{b}}$ and $k_{\mathrm{a}}$ are the spring constants
($4431.534$ kJ/mol/\AA$^{2}$ and $317.566$ kJ/mol/rad$^{2}$ respectively), $r_{ij}$ is the distance between atoms $i$ and $j$, $\varepsilon_{ij}$ and
$\sigma_{ij}$ are the Lennard-Jones parameters for atom pair $(i,j)$. In the electrostatic term, $q_{i}$ is the partial charge on atom $i$ (in units
of elementary charge $e$). In the SPC/Fw model, no Lennard-Jones
interactions are considered between oxygen and hydrogen, whereas oxygen-oxygen interaction parameters are  $\varepsilon_{\mathrm{OO}}=0.650$ kJ/mol and
$\sigma_{\mathrm{OO}}= 3.165$ \AA. Partial charges on oxygen and hydrogen atoms are $-0.82e$ and $+0.41e$ respectively.

%=======================================
\begin{figure}[b]
\includegraphics[width=8.5cm]{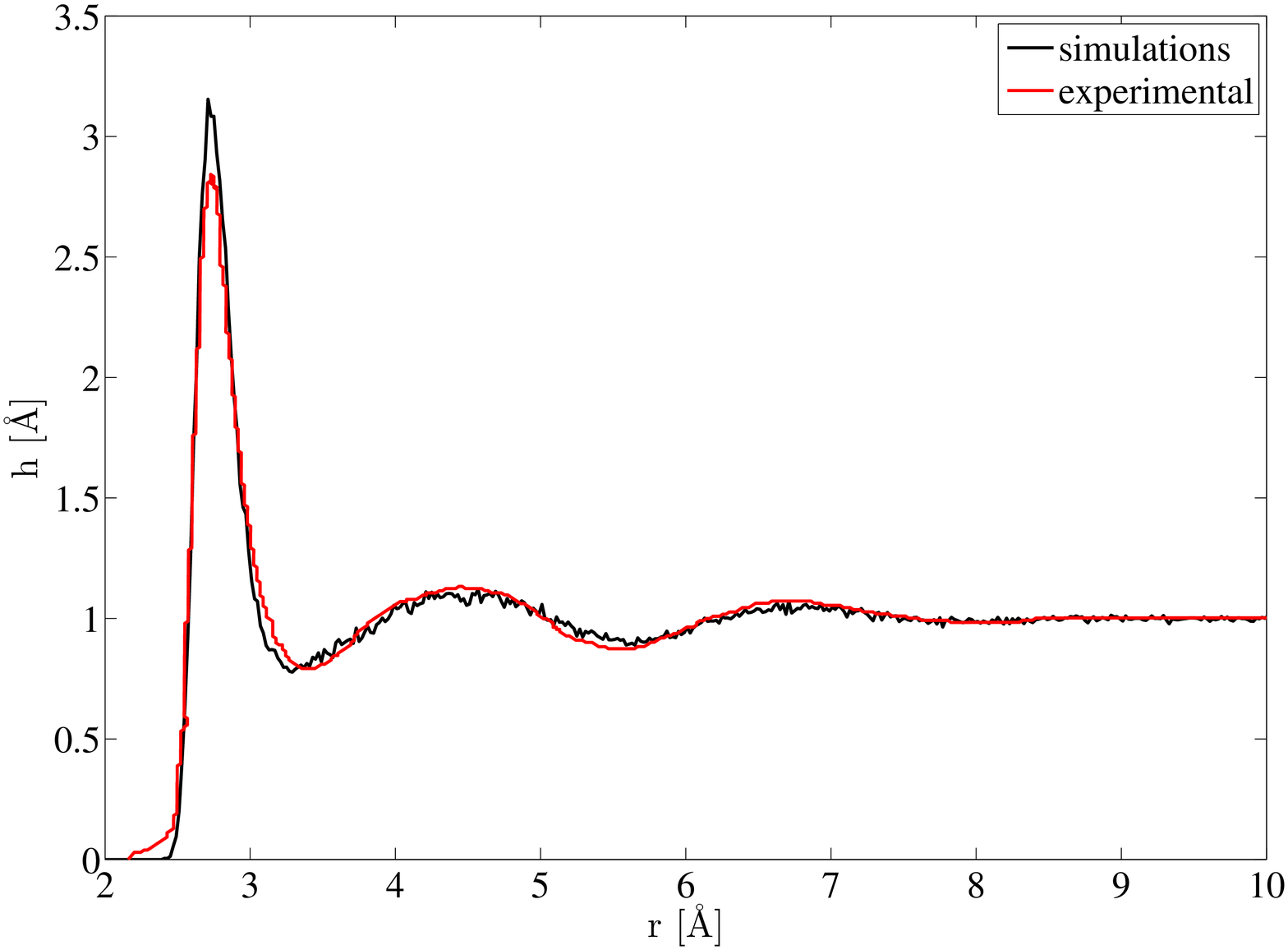}
\caption{
(Color online) Comparison between experimental and simulation data for oxygen-oxygen
radial distribution function. Experimental data are taken from Sorenson \textit{et al}.~\cite{rdf}.
\label{fig:rdf}}
\end{figure}
%=======================================

In our simulations we deviated from the original work describing the SPC/Fw model and chose to use a different set of parameters for the evaluation of
non-bond forces and for temperature and pressure control, more similar to those used in the works by Koumoutsakos and co-workers \cite{18,24,25}.
Temperature and pressure control were achieved using Berendsen thermostat and barostat \cite{49} while van der Waals and electrostatic forces were
treated using group-based cutoffs. Cutoff length was set at $r_{\mathrm{out}}=10$ \AA, with the potential smoothed to zero using a switching function
$S(r)$ along a width of $1$ \AA$^{}$ \cite{charmm}:
\begin{equation}
S(r)=\frac{(r^{2}_{\mathrm{out}}-r^{2})^{2}(r^{2}_{\mathrm{out}}+2r^{2}-3r_{\mathrm{in}}^{2})}{(r_{\mathrm{out}}^{2}-r_{\mathrm{in}}^{2})^{3}}\ ,
\label{eq:switch}
\end{equation}
where $r_{\mathrm{in}}=9$ \AA.

Because of this choice of parameter set, it was necessary to check if our scheme could give reliable results in the calculation of bulk water properties.
Therefore, we built a $4'000$ SPC/Fw water molecules box with periodic boundary conditions and we simulated its dynamical behavior in the NPT ensemble
for $6.5$ ns at a temperature of $298.15$ K and a pressure of $1$ atm, similar to what had been done in the original work \cite{43}. We restricted
ourselves to the calculation of basic properties, i.e. average bond length $\langle r_{\mathrm{OH}}\rangle$ and angle $\langle\theta_{\angle\mathrm{HOH}}\rangle$,
density $\rho$, oxygen-oxygen radial distribution function $g(r_{\mathrm{OO}})$ and self-diffusion constant $D_{\mathrm{s}}$. Results obtained for these analyses are
reported in Table \ref{tab:bulk}, along with comparisons with experimental measurements and results by Wu \textit{et al}.~\cite{43}. A graph of $g(r_{\mathrm{OO}})$ (experimental
and simulated) is reported in Figure \ref{fig:rdf}. The slight differences between measured quantities and referenced ones allowed us to regard our parameter set
as being effective for the simulation of bulk liquid water at that specific temperature. Moreover, we could not spot any of the artifacts that could
arise when using cutoffs for electrostatic terms in the simulation of water \cite{50,51}. Finally, as a final check for our settings, we calculated
surface tension $\gamma$ for the SPC/Fw water model. Starting from an average density frame extracted from the last ns of the NPT trajectory, we
performed an NVT simulation of a water slab by simply adding $50$ \AA$^{}$ of vacuum space in the $z$ direction. Dynamical behavior of the system was again
simulated for $6.5$ ns at a temperature of $298.15$ K, using a Berendsen thermostat and the previous cutoff scheme. Recording of pressure components
$P_{xx}$, $P_{yy}$ and $P_{zz}$ every $1$ ps during the last $5$ ns of the NVT simulation allowed us to calculate surface tension $\gamma$ by using the simple
formula \cite{52,53,54}:
\begin{equation}
\gamma=\frac{L_{z}}{2}\Big(P_{zz}-\frac{P_{xx}+P_{yy}}{2}\Big)\ ,
\end{equation}
where $L_{z}$ is the box length in the $z$ direction. Surface tension $\gamma$ estimated value was $70.8\pm 1.9$ mN/m, differing by only $1.7\%$ from
the experimental value of $72$ mN/m.
%=======================================
\begin{table*}[t]
\begin{center}
\begin{tabular}{c|c|c|c}
\hline
\hline
\multicolumn{2}{c}{spherical droplets} \vline & \multicolumn{2}{c}{cylindrical filaments}\\
\hline
radius $r_{\mathrm{i}}$ \AA & No.~water molecules & radius $r_{\mathrm{i}}$ \AA & No.~water molecules\\
\hline
$14.9$ & $250$ & $14.9$ & $322$\\
$18.6$ & $500$ & $18.6$ & $490$\\
$23.8$ & $1'000$ & $23.8$ & $737$\\
$30.1$ & $2'000$ & $30.1$ & $1'263$\\
$38.5$ & $4'000$ & $38.5$ & $2'068$\\
$48.1$ & $8'000$ & $48.1$ & $3'213$\\
$60.7$ & $16'000$ & $60.7$ & $5'117$\\
\hline
\hline
\end{tabular}
\end{center}
\caption{\label{tab:initial}
Initial radius $r_{\mathrm{i}}$ and number of water molecules of droplet and filament models.}
\end{table*}
%=======================================
%=======================================
\begin{figure}[b]
\includegraphics[width=4.77cm]{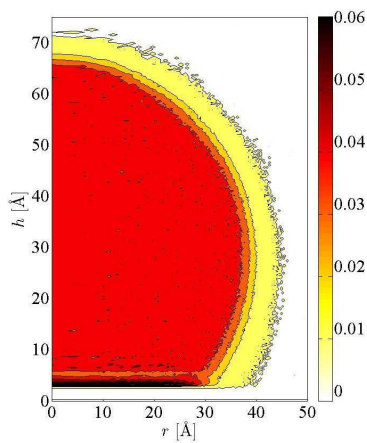}\hspace{0.5cm}
\caption{
(Color online)
Average density map for a spherical droplet comprising $8'000$ water molecules
and $\varepsilon_{\mathrm{CO}}=0.2$ kJ/mol. Color code based on density expressed as number of
molecules per \AA$^{3}$. Similar representations are obtained for the other droplets,
both spherical and cylindrical.
\label{fig:density}}
\end{figure}
%=======================================

\textit{Simulation of water on graphitic surface}. Once our parameter set was proven to hold reliable results for water bulk properties and surface tension,
we built models of water spherical droplets and cylindrical filaments, starting from an average density frame selected from the NPT simulation described
above. From this state, we extracted several hemispherical and hemicylindrical shapes, whose characteristics are reported in Table \ref{tab:initial}.
For the largest models, we simply replicated the $4'000$ molecule box in the three directions of space, in order to have enough  molecules for
our structures.

Hemispherical and hemicylindrical shapes, containing a number of water molecules ranging from $250$ to $16'000$, were then positioned just above
(approximately $2-3$ \AA) the model of  two graphene sheets arranged as hexagonal graphite, using periodic boundary conditions. We used
only two graphene sheets as long as additional ones would have been beyond the cutoff distance for any water molecule on the surface; moreover, graphene
sheet atoms were kept fixed during all simulations, consistently to what had been reported in previous works \cite{18}. Side lengths of the
graphene sheets
were varied according to the droplet shape and size. For spherical droplets, square planes were used, with sizes approximately ranging from
$100\times 100$
to $160\times 160$ \AA$^{2}$. For cylindrical droplets, we used rectangular planes, with a fixed thickness of $25.6$ \AA$^{}$ and a length ranging from
$100$ to $160$ \AA. The length of the simulation box in the $z$ direction was kept equal to the longest edge in the $xy$ plane. Box boundaries
were distant enough to prevent any interaction between periodic images for any of the simulated models.

All of the systems described above were simulated using different values of $\varepsilon_{\mathrm{CO}}$.
Simulation procedure was similar to the one used by Werder \textit{et al}.~\cite{18} and consisted of three different steps. Firstly, a simple
energy minimization was performed; secondly, for equilibration purposes, each system was simulated in the NVT ensemble for $0.5$ ns at a temperature
of $298.15$ K, using a timestep of $1$ fs. A Berendsen thermostat was used for temperature control and non-bond interactions were treated with the
scheme described above, i.e. with a cutoff distance $r_{\mathrm{out}}=10$ \AA$^{}$ and a smoothed potential calculated using Equation \ref{eq:switch}.
Finally, a $0.5$ ns
simulation was performed in the NVE ensemble, again with  a timestep of $1$ fs and the same non-bond interaction scheme.

%=======================================
\begin{figure}[b]
\includegraphics[width=3.5cm]{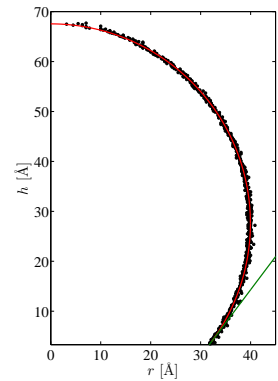}
\caption{\label{fig:profile}
(Color online) Equilibrium profile calculated with cylindrical binning procedure
(black points) and circular fit (red line) for a spherical droplet comprising
$8'000$ water molecules and $\varepsilon_{\mathrm{CO}}=0.2$ kJ/mol. Tangent line
determining contact angle is shown in green.}
\end{figure}
%=======================================
%=======================================
\begin{figure*}[t]
\includegraphics[height=5.5cm]{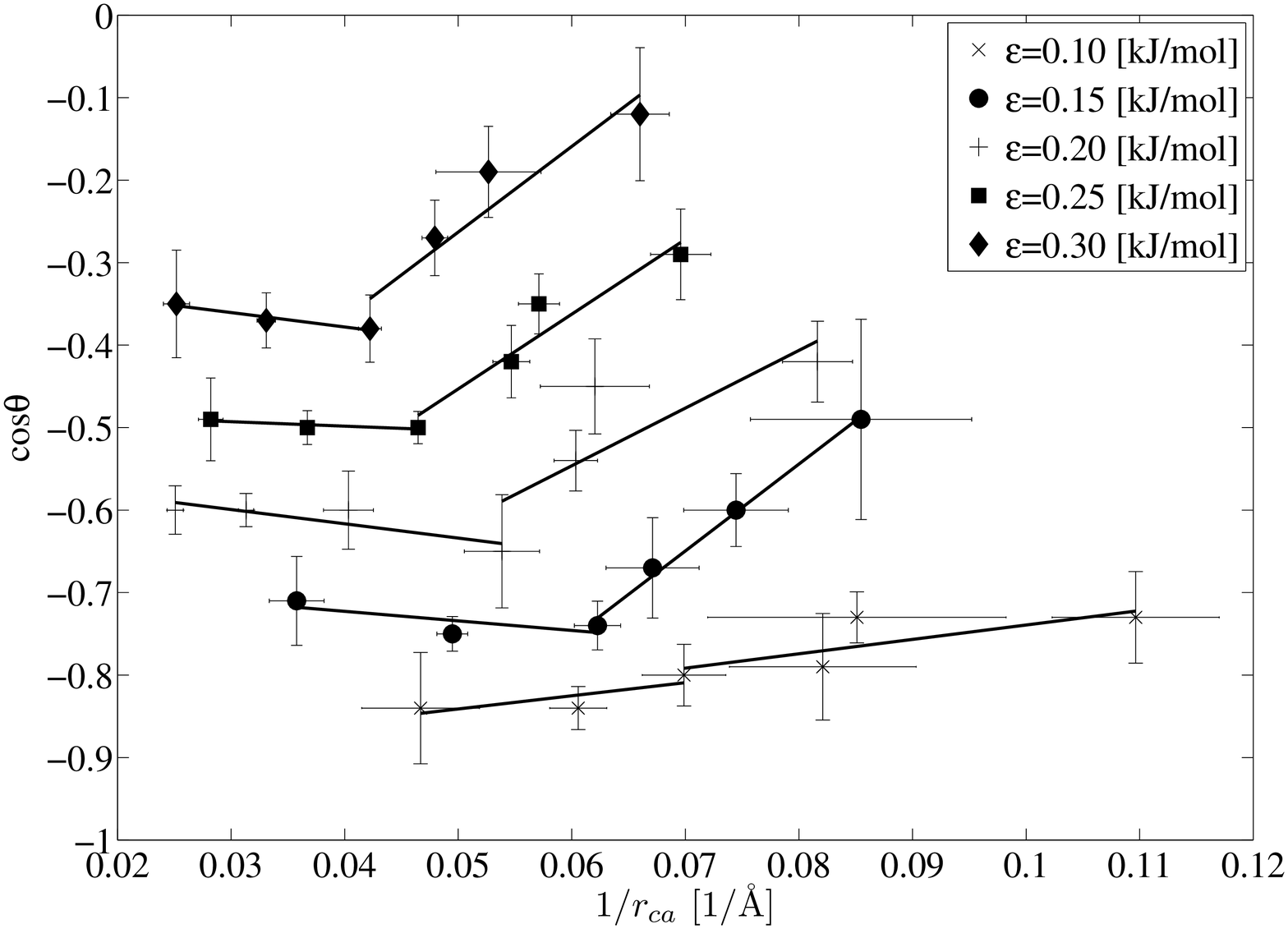}\hspace{0.5cm}
\includegraphics[height=5.5cm]{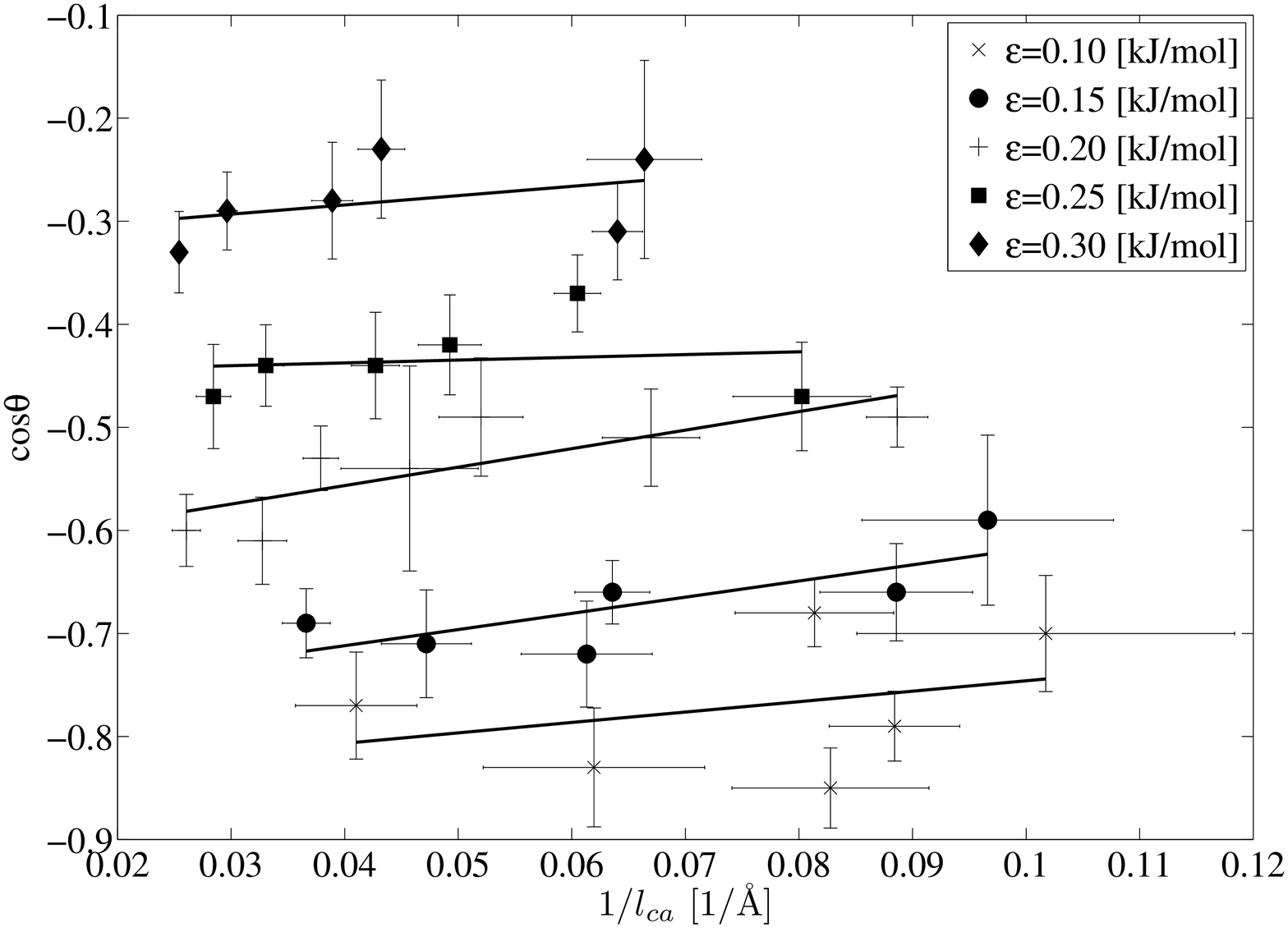}
\caption{
Left: Cosine of the contact angle $\theta$ as a function of $1/r_{\mathrm{ca}}$ for spherical
droplets. Right: Cosine of the contact angle $\theta$ as a function of $1/l_{\mathrm{ca}}$ for cylindrical
filaments. Error bars indicate standard deviation calculated by averaging over measurements taken on 5 blocks of 100 ps each during the NVE production run. 
\label{fig:regimes}}
\end{figure*}
%=======================================

\textit{Analysis.} In order to calculate the contact angle of spherical droplets and cylindrical filaments of water on graphene sheets, we first had to
define an equilibrium profile. For this purpose,  we used two variations of the analysis procedure reported by different authors \cite{18,35,36}.
For spherical droplets, the method basically consists of dividing the volume occupied by water molecules by using a cylindrical binning, regular
along the $z$ axis
and in the angle $\theta$. By imposing that the length of the $n$-th radial bin is $r_{n} = r_{0}\sqrt{n+1}$, all elements have the same volume
$\mathrm{d}V$, $r_{0}$ being the length of the first radial bin; in this  work, we chose $\mathrm{d}V = 0.009$ \AA$^{3}$. Reliability of this
choice was proven by considering different yet reasonable values for $\mathrm{d}V$ as well as for length and angle segmentation, which led to almost
identical final results for the contact angle measurement. It is worth noting that some care should be taken in order to avoid finite size effects.
For every frame of the NVE trajectory (frames were taken every $0.5$ ps), the number of atoms falling within a given volume element were counted and
averaged over both the total number of frames and the angle $\theta$, thus obtaining an equilibrium density map for the droplet
(Figure \ref{fig:density}). Water bulk
density $\rho_{\mathrm{b}}$ was calculated in the inner part of the droplet density map, i.e. for those volume elements located from $h_{\mathrm{a}}/4$ to $3h_{\mathrm{a}}/4$
along the $z$ axis and from $0$ to $r_{\mathrm{f}}/4$ along the radial axis ($h_{\mathrm{a}}$ and $r_{\mathrm{f}}$ being the approximate droplet height and
radius obtained from the density map).  For every layer in the $z$ axis, then, we detected those volume elements for which density first exceeded
$\rho_{\mathrm{b}}/2$, starting
our check from the vapor phase. In this way, we obtained equilibrium profiles which were subsequently approximated by a circular fit
(Figure \ref{fig:profile}).
Conversely to what had been done by other authors \cite{18,35,36}, we decided to include all points in the fitting in order not to introduce any arbitrary
cutoff for the definition of the circular profile. We also decided to use a circular fit in order to keep the whole procedure as simple as possible,
even though other investigators preferred to employ more sophisticated fits \cite{53}. Here we restrict ourselves to a circular fit because experimental
measurements are generally carried out under the assumption of a spherical shape of droplets.

For cylindrical filaments, we adopted an analogous procedure except for the fact that in this case averaging was done over the whole thickness of
the periodic box instead than over the angle $\theta$. In any case, we kept $\mathrm{d}V = 0.009$ \AA$^{3}$ also for these systems, in order to
allow a comparison as rigorous as possible between data obtained from cylindrical filament and spherical droplet simulations.
%======================================================================
%=======================================
\begin{table}[b]
\begin{center}
\begin{tabular}{c|c|c|c}
\hline
\hline
\multirow{2}{*}{$\varepsilon_{\mathrm{CO}}$ kJ/mol}& \multicolumn{2}{c|}{$\theta_{\infty}$} & \multirow{2}{*}{$\Delta\theta_{\infty}$} \\
\cline{2-3} 
 & spheres & filaments \\
\hline
0.10 & 157.0$^{\circ}\pm 6.9^{\circ}$ & 147.9$^{\circ}\pm 5.2^{\circ}$ & $9.1^{\circ}\pm 12.1^{\circ}$ \\
0.15 & 132.6$^{\circ}\pm 4.5^{\circ}$ & 140.8$^{\circ}\pm 6.4^{\circ}$ & $8.2^{\circ}\pm 10.9^{\circ}$\\
0.20 & 123.2$^{\circ}\pm 5.0^{\circ}$ & 128.9$^{\circ}\pm 6.8^{\circ}$ & $5.7^{\circ}\pm 11.8^{\circ}$\\
0.25 & 118.5$^{\circ}\pm 3.3^{\circ}$ & 116.6$^{\circ}\pm 3.4^{\circ}$ & $1.9^{\circ}\pm 6.7^{\circ}$\\
0.30 & 108.0$^{\circ}\pm 4.0^{\circ}$ & 108.7$^{\circ}\pm 5.7^{\circ}$ & $0.7^{\circ}\pm 9.7^{\circ}$\\
\hline
\hline
\end{tabular}
\end{center}
\caption{\label{tab:theta}
Extrapolated values of $\theta_{\infty}$ for different values of $\varepsilon_{\mathrm{CO}}$.}
\end{table}
%=======================================
%=======================================
\begin{figure*}[t]
\includegraphics[height=5.5cm]{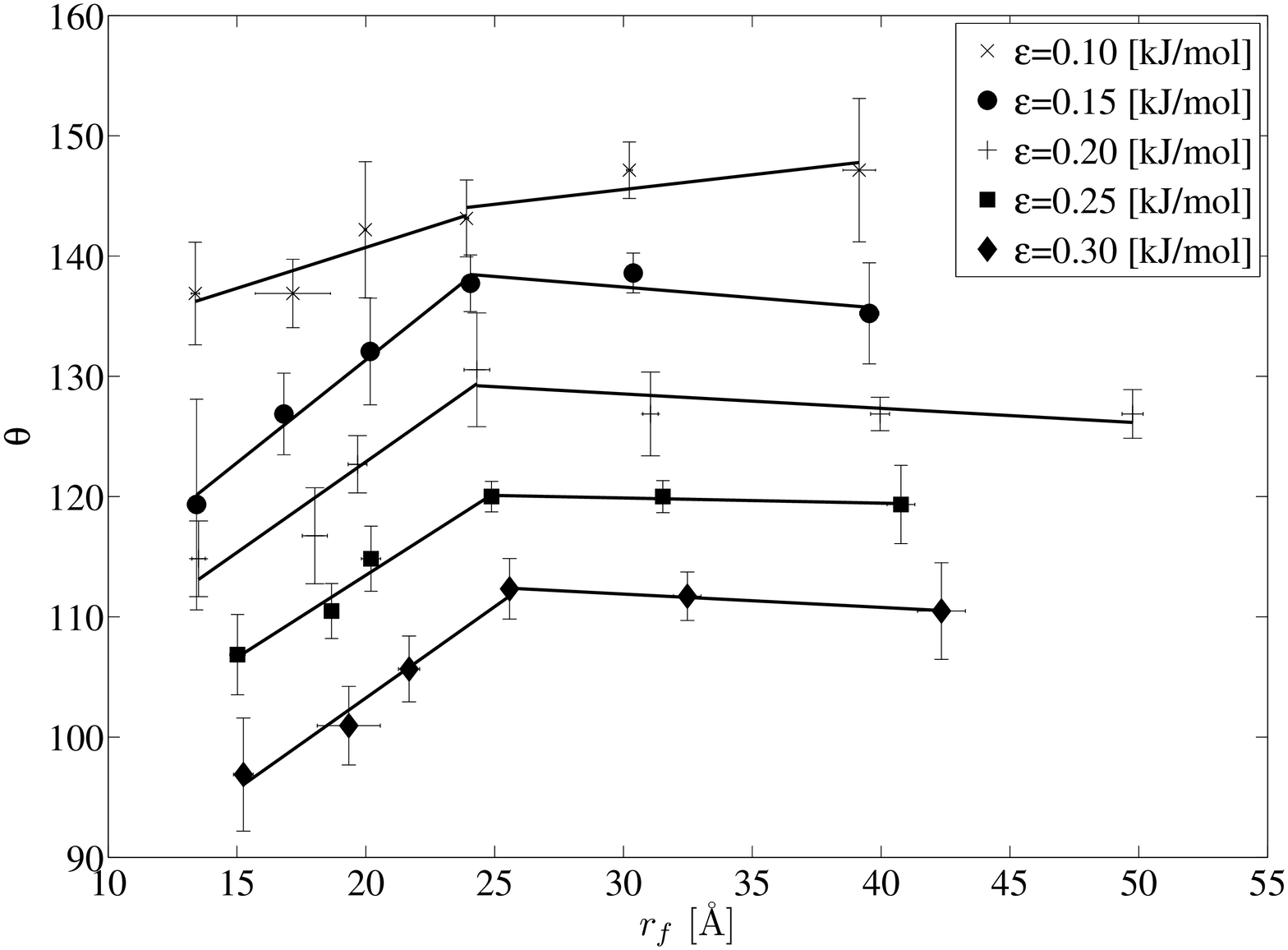}\hspace{0.5cm}
\includegraphics[height=5.5cm]{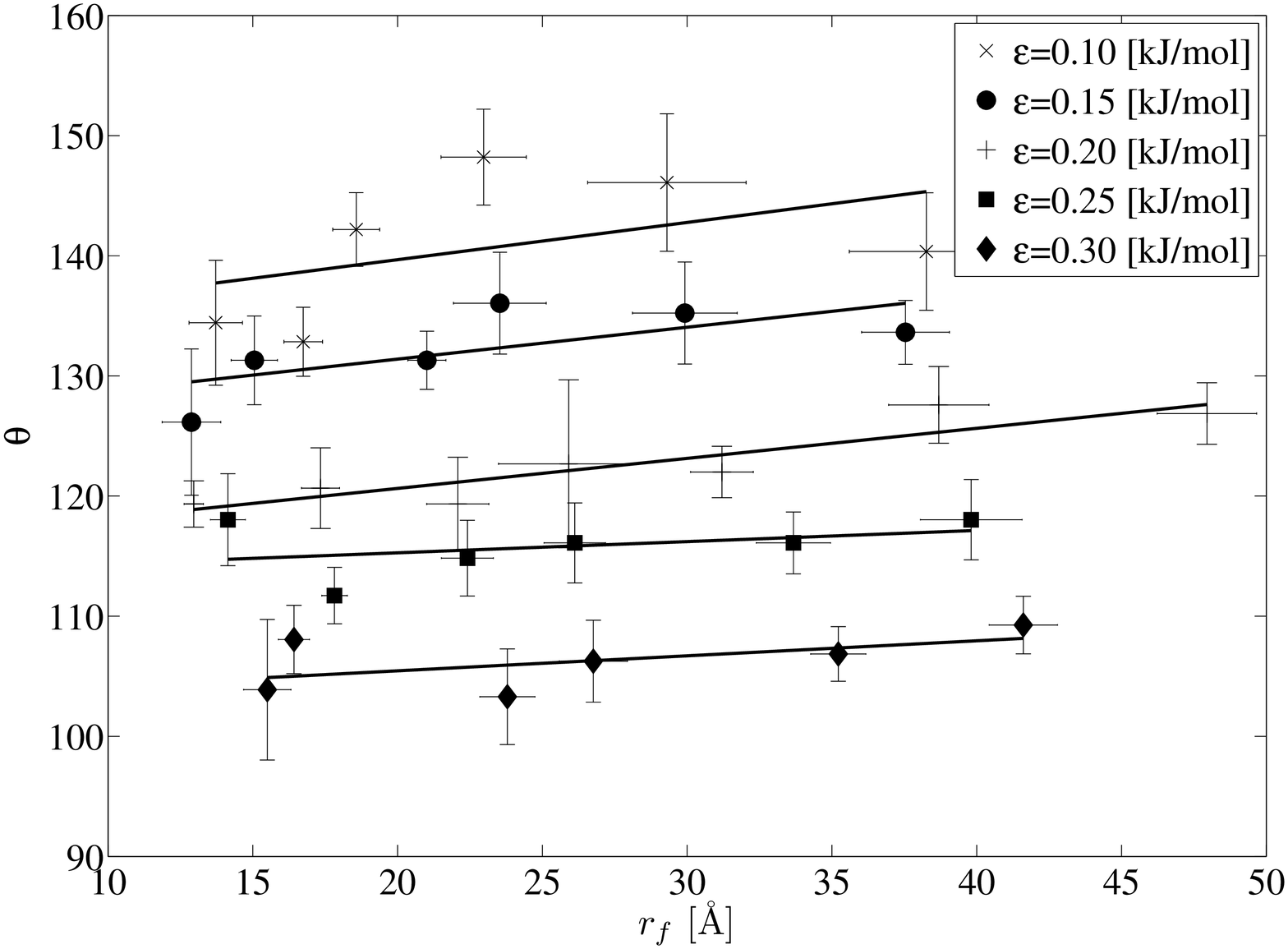}
\caption{
Contact angle $\theta$ as a function of $r_{\mathrm{f}}$ for spherical droplets, Left, and cylindrical filaments, Right. Error bars indicate standard deviation 
calculated by averaging over measurements taken on 5 blocks of 100 ps each during the NVE production run.
\label{fig:theta}}
\end{figure*}
%=======================================
%=======================================
\begin{figure}[t]
\includegraphics[width=8.5cm]{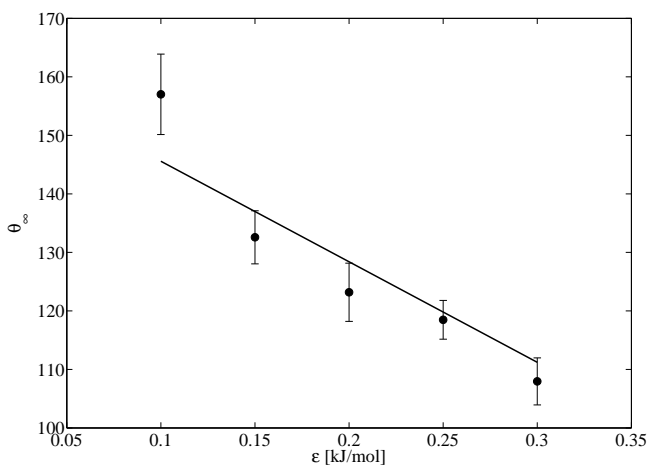}
\caption{
$\theta_{\infty}$ as a function of $\varepsilon_{\mathrm{CO}}$ for spherical droplets (cf.~Table
\ref{tab:theta}).
\label{fig:infty}}
\end{figure}
%=======================================

\section{Results and discussion}

In Figure \ref{fig:regimes}, cosines of contact angles $\theta$ calculated for spherical droplets are plotted against contact line curvature
$1/r_{\mathrm{ca}}$. A similar plot for cylindrical filaments is also reported, where $l_{\mathrm{ca}}$  is intended as the inverse of half the length
of the chord determined on a vertical section by the points at the intersection between the circular fit and the plane (see also Figure
\ref{fig:setup}). By plotting cosines of contact angles $\theta$ against these two quantities, we intend to provide a term of comparison
between droplets and filaments with the same $\varepsilon_{\mathrm{CO}}$ and initial radius.

From Equation \ref{eq:young}, we would expect a monotonous trend in the curves of Figure \ref{fig:regimes}, but for spherical droplets two different regimes
seem to be present. For larger droplets, contact angle slightly decrease with increasing $r_{\mathrm{ca}}$ and it is possible to determine an 
average positive line tension of $(5.060\pm 1.061)\cdot 10^{-12}$ N, in line with results reported in literature \cite{18,34}. 
Conversely, for very small droplets ($r_{\mathrm{i}}<30$ \AA), contact angle seems to decrease with decreasing $r_{\mathrm{ca}}$. 
In any case, if we consider only those droplets with
$r_{\mathrm{i}}>30$ \AA, we can easily extrapolate the macroscopic contact angle $\theta_{\infty}$ for the different $\varepsilon_{\mathrm{CO}}$ values
we considered (see Table \ref{tab:theta}). In order to match the experimental value of $127^{\circ}$, an appropriate estimation for this parameter for
SPC/Fw water should be around $0.2$ kJ/mol (see Table \ref{tab:theta} and Figure \ref{fig:infty}). This is significantly lower than what
is commonly used in oxygen-carbon Lennard-Jones potentials for modeling the interaction between water and graphitic surfaces
\cite{17,18,19,20,21,22,23,24,25,26,27,28,29}, assuming water is modeled with a set of parameters similar to the one described above (i.e. three point
charge models with a Lennard-Jones potential between oxygen atoms). 

%=======================================
\begin{figure}[t]
\includegraphics[width=8.5cm]{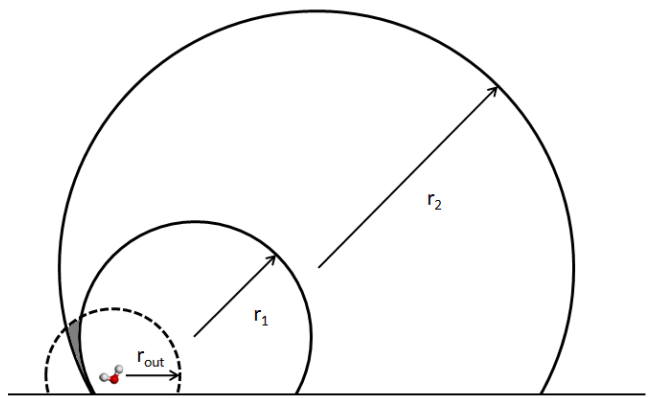}
\caption{
(Color online) Schematic representation of the section of two spherical droplets/cylindrical filaments 
with different radii $r_{\mathrm{1}}$ and $r_{\mathrm{2}}$ (section is intended perpendicular to the graphitic substrate and to the cylinder axis).  The dotted line 
represents the interaction range of a water molecule close to contact line. The two droplets/filaments can not possibly present the same value for the contact angle 
(as is depicted), not only because of line tension effects related to contact line curvature, but also because the resultant force acting on the molecule 
in the smaller droplet/filament lacks the contribute of the water molecules in the shaded area. 
\label{fig:range}}
\end{figure}
%=======================================

For cylindrical filaments, conversely, the value of contact angle should be independent from their size, as long as contact line curvature is zero for this geometry 
and therefore no contribute should be present from the term related to line tension in Equation \ref{eq:young}.
Nonetheless, we can see a slight increase in $\theta$ with $l_{\mathrm{ca}}$ (see Figure \ref{fig:regimes}), similarly to what is noticed for smaller droplets. 
Table \ref{tab:theta} reports contact angles for filaments if this trend is extrapolated to the macroscopic limit; as it would be expected, results show good agreement 
with the corresponding contact angles calculated for spherical droplets. 
Interstingly, if we plot contact angle versus $r_{\mathrm{f}}$ (i.e. the final radius of the sphere/cylinder) for both filaments and droplets
(see Figure \ref{fig:theta}), it is possible to see that $\theta$ in small droplets it is more sensitive to size variations via $r_{\mathrm{f}}$ 
than its counterpart in cylindrical filaments. The reason for this unexpected behavior in our MD simulations for both filaments and small droplets 
(i.e.~decreasing $\theta$ for decreasing droplet/filament size, hence contact angle values not consistent with modified Young's law) could reside in the fact that, 
in these systems, molecules near the contact line experience 
reduced cohesive forces because of the higher curvature of the liquid/vapour dividing surface and this results in an increase in the adhesive component of the total
force acting on them. This effect can be easily understood from the schematic representation of Figure \ref{fig:range}, where the interaction range of the molecule 
is depicted as a circle of radius $r_{\mathrm{out}}$. This reduction in cohesive forces has an opposite effect to the one related to a positive 
line tension (which for spherical droplets should in principle increase contact angle with decreasing droplet size) and it seems to be dominant over the one 
related to contact line curvature $1/r_{\mathrm{ca}}$ for spherical droplets with $r_{\mathrm{i}}<30$ \AA. A possible validation for this supposition comes from 
the fact that for cylindrical filaments, where no effect of line tension is expected, the increase in contact angle with increasing size is still present and 
it is less marked than for equivalent small spherical droplets, as it would be logical considering the difference of these surfaces near the contact line. Moreover,
it should be considered that as the droplet/filament grows smaller, the number of molecules which belong to the most superficial layer of the droplet increases 
relatively to those in the bulk. This could again result in the reduction of cohesive forces for molecules near the contact line and in the decrease of the 
contact angle. One could argue that these results are merely an artifact caused by the choice of using group or atom based cutoffs, but additional simulations using 
larger cutoffs (12.5 and 15 \AA) for  $\varepsilon_{\mathrm{CO}}=0.2$ kJ/mol gave similar results (data not shown), as already reported by other authors for 
atom based cutoffs \cite{25}, even if in that case convergence of contact angle measurements was reached for slightly larger values of $r_{\mathrm{out}}$. Similarly, 
it has been proven that more accurate methods for the treatment of long range interactions in this kind of MD simulations do not have a significant influence on 
the value of contact angle \cite{24}. Finally, it is worth noting that both spherical droplets and cylindrical filaments, simulations with the lowest value of
$\varepsilon_{\mathrm{CO}}$ correspond to more scattered results and deviations from the behavior noted for all other values of the interaction parameter 
(see Figures \ref{fig:regimes}, \ref{fig:theta}, \ref{fig:infty}), indicating that when adhesive forces are extremely low it becomes more difficult to unambiguously
define and measure contact angle for the simulated systems.                        
%======================================================================

\section{Conclusions}

We have run MD simulations of spherical droplets and cylindrical filaments of SPC/Fw water on two sheets of graphene arranged as hexagonal graphite and
calculated the relative equilibrium contact angles along the contact line. Using system parameters similar to those previously adopted 
in the literature for the simulation of spherical droplets, results indicate that, in order to 
recover the recently measured macroscopic contact angle of $127^{\circ}$ between water and graphene, the value of $\varepsilon_{\mathrm{CO}}$ 
in the Lennard-Jones potential used to describe non bonded interactions between oxygen and carbon should be around $0.2$
kJ/mol, much lower than the one that is commonly used in MD studies. Comparison between contact angle measurements for spherical droplets and cylindrical filaments
seems to indicate that these two systems lead to comparable results in the macroscopic limit, but both systems show that contact angle varies with drop size 
in a way which is not consistent with modified Young's equation. As already suggested by other authors \cite{SD}, it seems that, beyond line tension correction, 
other contributions should be considered in modifying Young's equation for extremely small droplets/filaments; we believe these contributions could be related 
to the dividing surface curvature and to the relative increase in the number of molecules belonging to the most superficial layer 
with respect to the bulk phase. In any case, it has to be noticed that these results have been obtained in a very specific framework and, to a certain extent, are
dependant on the choice of the parameters used in the simulation. Moreover, due to the challenges presented by measurements of contact angle for such small 
systems, it would be very difficult to seek an experimental confirmation to the reported behavior. Therefore, we believe that further work in 
MD simulation of this sort is necessary, in order to sistematically investigate the influence of all different parameters and thus to help in the 
understanding of wetting phenomena at the nanoscale. 

%======================================================================

\begin{acknowledgments}
This work was supported by the Swiss Commission for Technological Innovation (CTI) through project BiPCaNP (CTI P.~No.~10055.1).
\end{acknowledgments}

%===========================================================================

%=======================================================================
\end{document}